# Modified Mott-Schottky Analysis of Nanocrystal Solar Cells

**S. M. Willis, C. Cheng, H. E. Assender and A. A. R. Watt**
*Department of Materials, University of Oxford, Parks Road, Oxford. OX1 3PH. United Kingdom*

Mott-Schottky analysis is adapted to determine the built-in bias ($V_{bi}$) and doping density ($N$) of lead sulfide-zinc oxide colloidal quantum dot heterojunction solar cells. We show that charge injection barriers at the solar cell's electrodes create a constant capacitance that distorts the junction's depletion capacitance and result in erroneous $Vbi$ and $N$ values when determined through Mott-Schottky analysis. The injection barrier capacitance is taken into account by incorporating a constant capacitance in parallel with the depletion capacitance.

Mott-Schottky analysis is commonly used to determine the built-in bias ($V_{bi}$) and doping density ($N$) of a semiconductor at Schottky and *p-n* junctions[1]. However, the method is not always straightforward. For example, when the analysis is applied to organic solar cells the parameters are frequency dependent, making it difficult to extract the true values[2]. Furthermore, sources of capacitance other than the depletion capacitance distort the capacitance-voltage (CV) response, resulting in values of $V_{bi}$ and $N$ that are larger than the actual values. In this paper we present a capacitance model to explain the higher values and apply it to a PbS/ZnO colloidal quantum dot (CQD) heterojunction solar cell.

Mott-Schottky analysis probes the depletion capacitance at a Schottky or *p-n* junction which is determined by the width of the bias-dependent depletion region. Hence the depletion capacitance, $C$, is also bias dependent and can be expressed as[3,4]

$$\frac{1}{C^2} = \frac{2(V_{bi} - V)}{A^2 q \varepsilon \varepsilon_0 N} \qquad (1)$$

where $V$ is the applied bias, $A$ is the device area, $q$ is the elementary charge, $\varepsilon$ is the material's dielectric constant, and $\varepsilon_0$ is the permittivity of free space. The built-in bias and doping density are then found by fitting equation 1 to the linear portion of the $C^{-2}$ versus bias voltage plot[5].

However, Mott-Schottky analysis on PbS/ZnO CQD solar cells that have been irradiated with UV light to photodope the ZnO (details in the supporting information) was found to give parameters larger than expected ($V_{bi}$ = 0.6 V, $N$ = 2.5 x $10^{17}$ cm$^{-3}$). The $V_{bi}$ is expected to be near 0.2 V[6,7], and the PbS doping density, $N_a$, was found to be near 4 x $10^{16}$ cm$^{-3}$ from Mott-Schottky analysis of PbS CQD/Al Schottky cells we fabricated. For cells containing heavily doped ZnO ($N$~$10^{19}$ cm$^{-3}$)[8], the depletion region should lie almost entirely in the PbS layer. Hence the heterojunction results should reflect the doping density of the PbS layer. The high values for the two parameters in the PbS/ZnO CQD cells can be explained by including a constant capacitance due to charge injection barriers at the electrodes as a part of the overall capacitance.

To account for injection barriers, the overall capacitance can be modeled as two capacitors in parallel[9] as shown in Figure 1a. Figure 1b displays a reference CV curve taken from a gallium arsenide *p-i-n* solar cell that was chosen because it has a well understood depletion capacitance. It also shows CV curves, derived from this reference curve with different values of constant capacitance added in parallel to the reference capacitance to create a simulated overall capacitance. Values of constant capacitance between 1 and 7 nF were chosen for the simulation since these values are on the same order as the geometric capacitance of the cell structure.

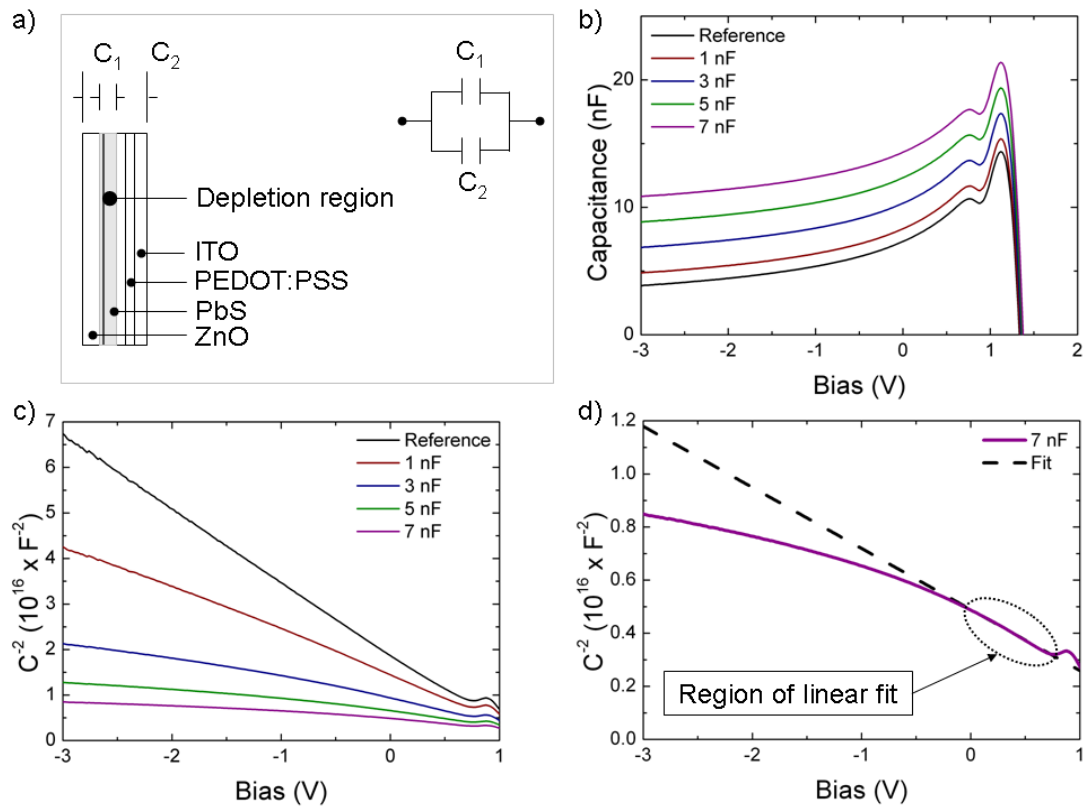

**Figure 1. a) PbS/ZnO solar cell with the origin of depletion capacitance, $C_1$, and a constant capacitance, $C_2$, due to injection barriers. The equivalent circuit is two capacitors in parallel. b) CV curves for the reference data and different values of constant capacitance in parallel. c) $C^{-2}$ versus bias curves for the reference data and different values of constant capacitance in parallel. d) Expanded view of the reference capacitance with 7 nF constant capacitance in parallel. Not only does the slope and x-axis crossing change, but the shape of the curve is distorted as well**

For larger values of the injection barrier capacitance, the $C^{-2}$ curve is seen to deviate more from the reference curve (Figure 1c). Mott-Schottky fits then produce larger $V_{bi}$ (*x*-axis crossing) and $N_A$ (inversely proportional to the slope) values than the reference data. The linear region of the $C^{-2}$ slope itself also becomes distorted as is illustrated in Figure 1d for the example of a constant 7 nF capacitance added in parallel to the reference capacitance. Linear regions still exist, however, that can be fitted with equation 1. The fitting results produce built-in bias and doping density values higher than the true value for the *p-i-n* junction. For the simulation here with a constant 7 nF capacitance due to injection barriers, the fit gives $V_{bi}$ = 2.1 V compared to the true value (no constant capacitance) of 1.1 V. Likewise, the fit gives $N_A$ = 7.6 x $10^{16}$ cm$^{-3}$ compared to the true value of 1.1 x $10^{16}$ cm$^{-3}$.

An example of this model applied to the capacitance of our PbS/ZnO cells is shown in Figure 2a-b. The PbS doping level was found to be ~4 x $10^{16}$ cm$^{-3}$ by taking CV measurements of a PbS/Al Schottky cell and fitting the $C^{-2}$ curve with equation 1. Similar values have been reported in literature[3]. Using a value of 2.7 nF for the constant capacitance due to the injection barrier gives a good fit of the adjusted depletion capacitance with fit parameters, close to those expected, of $V_{bi}$ = 0.24 V and $N_A$ = 4 x $10^{16}$ cm$^{-3}$ (Figure 2b). The calculated geometric capacitance based on the cell's dimensions is ~2 nF.

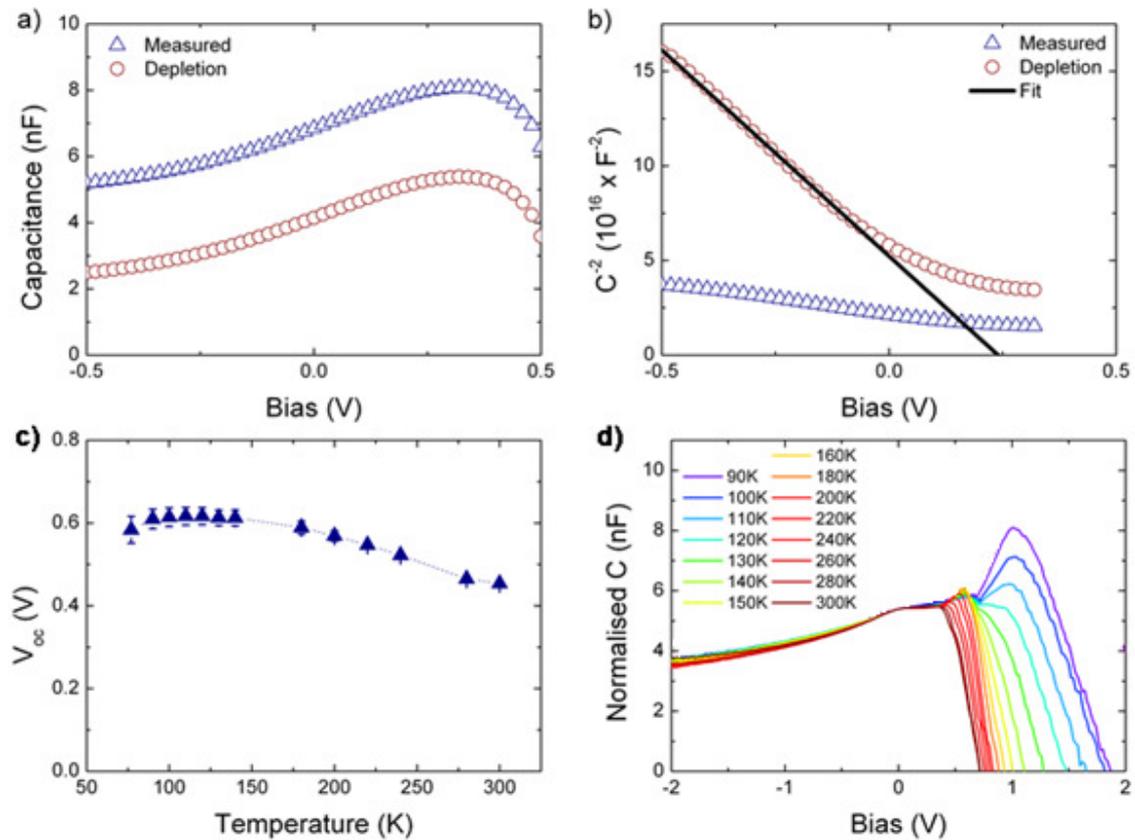

**Figure 2. a) Measured CV (blue triangles) at room temperature. The depletion capacitance (red circles) was calculated based on the parallel capacitors equivalent circuit shown in Figure 1a with a constant capacitance of 2.7 nF. b) The corresponding $C^{-2}$ curves with a Mott-Schottky fit to the depletion capacitance. $V_{bi}$ = 0.24 V, $N_A$ = 4 x $10^{16}$ cm$^{-3}$. c) Temperature-dependent $V_{oc}$. d) Temperature-dependent CV adjusted to coincide at 0 V bias. Below ~130 K a sharp rise in capacitance is observed before the rapid decrease.**

In the proposed capacitance model, the source of constant capacitance was based on charge injection barriers at the electrodes. Evidence for the barriers is seen in the temperature-dependent current-voltage (*JV*) and CV results. The temperature-dependent open circuit voltage ($V_{oc}$) is shown in Figure 2c. A near linear increase is seen with temperatures decreasing from 300 K to ~160 K. With further temperature decrease, the $V_{oc}$ saturates and begins to drop again. A similar temperature-dependent $V_{oc}$ behavior was observed in organic bulk heterojunction solar cells[10], which was modeled in terms of charge injection barriers at the electrodes[10, 11]. Likewise, charge injection barriers lead to the increased Schottky analysis parameters we propose here.

Figure 2d shows the bias-dependent capacitance measurements for temperatures in the range of 90 – 300 K. For clarity, the capacitance values have been adjusted so that the capacitance for each temperature is equal at 0 V applied bias; the unadjusted values are given in the supporting information. At lower temperatures there is a sudden rise in capacitance, and at all temperatures there is a peak in capacitance at the bias at which the capacitance begins a rapid decrease becoming inductive.

Excessive capacitance peaks beyond the space-charge capacitance similar to those observed here have also been reported in Schottky diodes for many decades and have been shown to be due to the injection of minority carriers[9,12]. Carrier injection has also been given as the reason for excessive capacitance peaks in polymer LEDs[13], silicon *p-n* solar cells[5], and organic bulk heterojunction solar cells[4]. For a perfectly Ohmic back contact, the minority current has an inductive effect in Schottky diodes resulting in a reduced or even negative capacitance since the minority current through the bulk is controlled by diffusion and flows out of phase with the applied voltage[9]. However, when high injection barriers are present, the capacitance is found to first produce an increased capacitance peak before becoming inductive[12].

Above ~130 K the sharp rise in capacitance is not seen and the onset of the drop in capacitance is located at the same bias as the diode turn-on voltage (Figure 3b). This corresponds to the start of electron transfer from the ZnO to the PbS where the effect of minority carrier injection into the PbS has an inductive effect on the overall capacitance since the electrons are either extracted at the PEDOT:PSS/PbS interface or combine with majority holes. The large rise in capacitance is seen at temperatures below ~130 K and the start of the decrease in capacitance no longer corresponds to the diode turn-on voltage (Figure 3a), demonstrating the effect of charge injection barriers: at the increase in capacitance, holes are injected into the PbS which increases the amount of charge and hence the capacitance. Only at greater bias can the additional charges be extracted, and the capacitance falls once more. Since the drop in capacitance occurs at biases beyond the diode turn-on voltage (Figure 3a), once this bias is reached, electrons injected into the ZnO can pass easily into the PbS layer to be extracted at the PEDOT:PSS or combine with holes causing the decrease in capacitance.

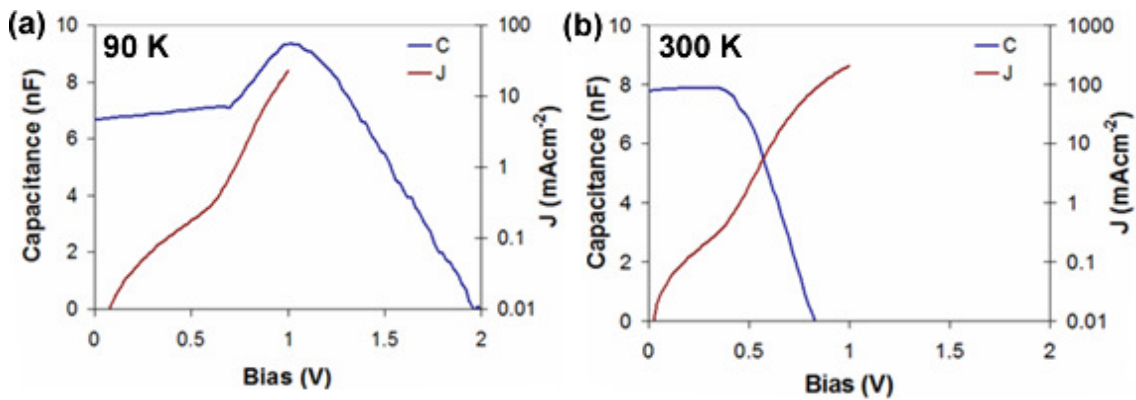

**Figure 3.** CV and JV comparison at low temperature (a) and high temperature (b).

In summary, charge injection barriers at a solar cell's electrodes can create a constant capacitance that distorts the junction's depletion capacitance and result in erroneous $V_{bi}$ and $N$ values when determined through Mott-Schottky analysis. The injection barrier capacitance can be taken into account by incorporating a constant capacitance in parallel with the depletion capacitance. Using this method, the true parameter values can be recovered.

---

[1] D. K. Schroder, Semiconductor Material and Device Characterization, Third Edition. Wiley – IEEE Press (2006)

[2] G. Jarosz, J. Non-Cryst. Solids. 354, 4338 (2008)

[3] J. P. Clifford, K.W. Johnston, L. Levina, E. H. Sargent, Appl. Phys. Lett. 91, 253117 (2007)

[4] G. Garcia-Belmonte, A. Munar, E.M. Barea, J. Bisquert, I. Ugarte, R. Pacios, Organic Electronics. 9, 847 (2008).

[5] I. Mora-Sero, G. Garcia-Belmonte, P.P. Boix, M.A. Vazquez, J. Bisquert, Energy & Environ. Sci. 2, 678 (2009)

[6] A. G. Pattantyus-Abraham, I.J. Kramer, A.R. Barkhouse, X. Wang, G Konstantatos, R. Debnath, L. Levina, I. Raabe, M. Nazeeruddin, M. Gratzel, E. H. Sargent, ACS Nano. 4, 3374 (2010)

[7] J. Gao, J.M Luther, O. E. Semonin, R. J. Ellingson, A.J. Nozik, M. C. Beard, Nano Lett. 11, 1002 (2011)

[8] G. Lakhwani, R. F. H. Roijmans, A. J. Kronemeijer, J Gilot, R.A.J Janssen, S. C. J Meskers, J. Phys. Chem. C. 114, 14804 (2010)

[9] J Werner, A. F. J. Levi, R.T. Tung, M. Anzlowar, M Pinto, Phys. Rev. Lett. 60, 53 (1988)

[10] D. Rauh, A. Wagenpfahl, C. Deibel, V. Dyakonov, Appl. Phys. Lett. 98, 133301 (2011)

[11] A. Wagenpfahl, C. Deibel, V. Dyakonov, IEEE Journal of Selected Topics in Quantum Electronics. 16, 1759 (2010)

[12] M. A. Green, J. ShewChun, Solid State Electron. 16, 1141 (1973)

[13] V. Shrotriya, Y. Yang, J. Appl. Phys. 97, 054504 (2005)